\documentclass[12pt]{article}

\usepackage[letterpaper,margin=1in]{geometry}
\usepackage[utf8]{inputenc}
\usepackage[T1]{fontenc}
\usepackage{setspace}
\usepackage{amsmath,amssymb}
\usepackage{bm}
\usepackage{graphicx}
\usepackage{subcaption}
\usepackage{xcolor}
\usepackage{authblk}
\usepackage[colorlinks=true,citecolor=blue,linkcolor=blue,urlcolor=blue]{hyperref}

\title{\bfseries Strong suppression of the photonic density of states in three-dimensional disordered silicon networks}

\author[1]{Abraham Aguilar Uribe}
\author[1]{Francisco Hern\'andez Alejandre}
\author[1]{Mattis Reisner}
\author[2,1]{Geoffroy Aubry}
\author[1]{Luis S.\ Froufe-P\'erez}
\author[3]{Marian Florescu}
\author[1,\dag]{Frank Scheffold}

\affil[1]{Department of Physics, University of Fribourg, Chemin du Mus\'ee 3, 1700 Fribourg, Switzerland}
\affil[2]{Universit\'e C\^ote d'Azur, CNRS, Institut de Physique de Nice -- INPHYNI, France}
\affil[3]{Optoelectronics Research Centre, University of Southampton, SO17 1BJ Southampton, UK}

\date{}

\begin{document}

\maketitle
\vspace{-2.5em}
\begin{center}
\dag\,Corresponding author. Email: Frank.Scheffold@unifr.ch
\end{center}
\vspace{1em}

\noindent\textbf{Abstract:} Photonic bandgaps can open not only in crystalline dielectric materials but also in amorphous structures. Self-uniform amorphous gyroid networks have been proposed as promising disordered photonic architectures. Motivated by numerical studies, we combine direct laser-writing lithography with advanced materials processing to fabricate these structures from high-refractive-index silicon. Spectroscopic measurements at mid-infrared wavelengths reveal pronounced transmission minima. To investigate the formation of photonic band gaps in amorphous structures, we develop polarization-resolved transmission spectroscopy that separates ballistic and multiply scattered contributions, enabling direct identification of the underlying transport regimes. We observe a strong suppression of diffuse transmission, providing direct experimental evidence for a breakdown of conventional diffuse transport associated with a strongly reduced photonic density of states. Together with large-scale numerical simulations, our results establish the presence of a deep photonic pseudogap in an amorphous three-dimensional dielectric material and open new opportunities for observing disorder-induced localization phenomena, including Anderson localization of light.

\bigskip

\section*{INTRODUCTION}

{\bf{D}}ielectric media can produce bandgaps for light in a manner analogous to the formation of electronic bandgaps in semiconductors~\cite{joannopoulos1997photonic,he2020colloidal}. Just as the de Broglie waves of electrons can interfere to prevent propagation within certain energy ranges, light waves can also interfere such that all waves are reflected within the bandgap frequency range, leading to evanescent attenuation of the light. The superposition of light waves can be treated classically, and the concept of Bloch waves in a periodically repeating environment allows the calculation of the photonic band structure.

Photonic band-gap materials---often described as semiconductors for light---have long attracted considerable interest as a platform for controlling and manipulating light, and as enabling technologies for optical communication and computing. They already play important roles in laser technology~\cite{wadsworth2002supercontinuum}, optical multiplexing for telecommunications~\cite{sahni2012silicon}, and solar energy harvesting~\cite{liu2019advance}, and, more recently, photonic-crystal architectures have also been explored for neuromorphic computing, where strong optical confinement and intrinsic nonlinearities can enable compact and energy-efficient information processing~\cite{Ji2025Neuromorphic}.

Amorphous electronic semiconductors have long been known to exist, demonstrating that a material can possess an electronic bandgap even in the absence of long-range crystalline order. This is possible because electrons can be localized around atomic cores, giving rise to the tight-binding description of band formation—a fundamental concept in solid-state physics that does not rely on periodicity~\cite{kittel2018introduction}. For photons, the situation is less obvious. Photonic bandgaps are traditionally understood as a consequence of Bragg interference in periodic dielectric structures, and therefore long-range order was long considered essential. However, recent theoretical and experimental work has provided evidence that disordered photonic materials may also support bandgaps. This progress has been driven by numerical studies of disordered diamond networks and stealthy hyperuniform structures, as well as experiments in two dimensions and in the microwave regime~\cite{man2013isotropic,muller2013silicon,yu2021engineered,siedentop2024stealthy}.

In the present work, we demonstrate a strong suppression of the photonic density of states in three-dimensional disordered dielectrics at mid-infrared wavelengths. Using direct laser writing and advanced materials processing, we fabricate silicon-based disordered networks based on the concept of local self-uniformity (LSU), resulting in amorphous gyroid structures. In a disordered medium, exponential attenuation of incident light is insufficient to identify a photonic band gap or pseudogap, since strong attenuation can also arise from scattering. Instead, we diagnose such a regime through the breakdown of classical diffuse transport, namely transmission falling below the diffusion bound, together with an independently computed suppression of the DOS. We introduce polarization-resolved transmission spectroscopy and a method to separate ballistic from multiply scattered light. Combined with large-scale numerical calculations, this enables us to identify a pronounced suppression of the photonic density of states in a regime where optical transport is no longer described by classical scattering theory or photon diffusion.

\section*{RESULTS}

\subsection*{Network design}
Different strategies for designing disordered dielectric materials with large photonic bandgaps have been proposed over the past years. All of these approaches share a common reliance on network topology and volume filling fractions in the range of 20--30\%~\cite{liew2011photonic}. Prominent examples include amorphous diamond structures~\cite{edagawa2008photonic}, stealthy hyperuniform networks~\cite{florescu2009designer,liew2011photonic}, dry foams~\cite{ricouvier2019foam}, and locally self-uniform gyroid networks~\cite{sellers2017local}. These dielectric architectures are predicted to exhibit full photonic bandgaps, which typically emerge when the refractive index contrast exceeds a threshold of $n/n_h \gtrsim 2$, where $n$ is the refractive index of the material and $n_h$ that of the host medium~\cite{HO1994413}. The gap width and threshold values are generally comparable to those observed in their crystalline counterparts, such as crystalline diamond or gyroid networks~\cite{soukoulis2011past,sellers2017local}.

In the present work, we study amorphous gyroid networks (AGN) proposed by Sellers et al.~\cite{sellers2017local}, which have been shown to be among the most effective disordered structures to open a full photonic bandgap. AGNs are prime examples of local self-uniform structures, where local self-uniformity (LSU) is a continuous metric bounded between 0 and 1 that quantifies the degree to which a network exhibits strong isotropy \cite{sellers2017local,Sunada}. The states of maximal LSU, corresponding to the upper bound of 1, are uniquely realised in three dimensions by the diamond and single gyroid topologies. More generally, the LSU framework evaluates the similarity of local motifs across a network, thereby enabling the systematic classification of disordered architectures of fixed connectivity along a continuum ranging from crystalline order to complete randomness. In particular, in the context of photonic band gap structures, these networks exhibit a local three-fold node topology---rather than the more common four-fold connected networks, typical for diamond- and diamond-like networks in tetrahedral geometries. AGNs also feature longer rods connecting the nodes, resulting in a more open structure that is particularly advantageous for the fabrication process we employ. The trivalent topology of LSU structures substantially simplifies fabrication compared to conventional four-connected photonic crystal networks, enabling robust architectures with smaller feature sizes or higher fidelity. To construct amorphous gyroid networks (AGNs)~\cite{sellers2017local}, we employ a modified Wooten--Winer--Weaire annealing protocol~\cite{Wooten1985}, analogous to the established approach to generate amorphous silicon models~\cite{Mousseau2002}, but incorporating a tailored effective functional ``potential energy'' that biases local coordination towards gyroidal arrangements.
The AGNs used here, a representative realization of which is shown in Fig.~\ref{fig:Fig1}, have pronounced local self-uniformity, with LSU metric values~\cite{sellers2017local} of 0.99 for single-vertex and 0.89 for two-vertex local configurations.

\subsection*{Fabrication}

\begin{figure}[p]
    \centering
    \includegraphics[width=\textwidth]{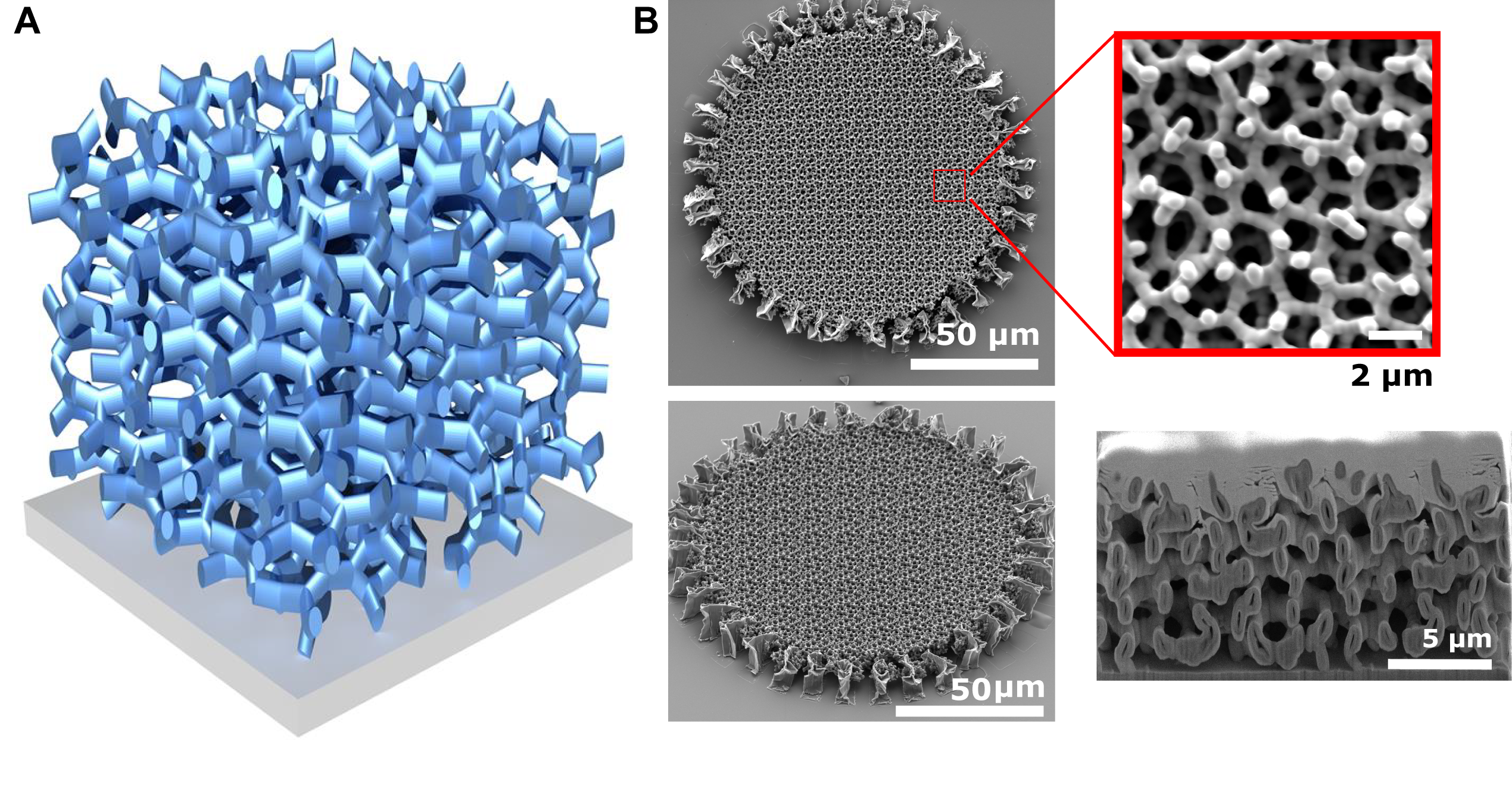}
\caption{ \textbf{Fabrication of silicon amorphous gyroid networks. }
\textbf{(A)} Digital rendering of an amorphous gyroid network designed following the approach described in~\cite{sellers2017local}. The average length of the rods connecting the nodes is $\bar{d}_0 = 0.8\,\mu$m. The rods are elliptical with an aspect ratio of 2.5 between the long and short axes.
\textbf{(B)} SEM images of a structure with thickness $L=11.6\mu$m after silicon infiltration and calcination: top view and inclined view. Enlarged view (scale bar: $2~\mu$m). A focused ion beam cut of the sample reveals the sample interior (side view). Contrast was enhanced using CLAHE (clip limit $=2.0$, tile grid size $=8 \times 8$)~\cite{zuiderveld1994contrast}.
}    \label{fig:Fig1}
\end{figure}

We prepare a large digital template by tiling previously
designed cubes of AGN with edge length $11.44\,\mu\mathrm{m}$, respecting periodic boundary conditions, Fig.~\ref{fig:Fig1}~A. The supercells are sufficiently large to suppress finite-size and superlattice effects over the spectral range of interest. The average rod (or cylinder) length is $\bar d_0 = 0.8~\mu\mathrm{m}$. After fabricating the rods with direct laser writing (DLW) lithography, they are elliptical, with an aspect ratio of $\sim$2.5 and the long axis oriented vertically~\cite{Aeby2021}. To adapt the digital designs to a stable and well-anchored writing volume, we position the digital template interface \(0.3~\mu\mathrm{m}\) inside the CaF$_2$ substrate, as is commonly done to ensure firm attachment. We also remove unconnected rods along the axial directions, which reduces the material density near the air interfaces. We determine the effective thickness by fitting the axial density profile of the digital template with an error function, finding a cumulative thickness reduction of about \(0.8~\mu\mathrm{m}\).
The total thickness is therefore given as $L \simeq \big(L_\text{nominal} - 0.8~\mu\mathrm{m}\big),$ and these corrected values are used throughout. We coat the polymer template with a roughly 10~nm thick layer of titanium dioxide at moderate temperatures (\(\sim 130^\circ\mathrm{C}\)). The titanium dioxide increases the volume filling fraction by approximately 1\% and stabilizes the structure against further shrinkage and structural damage. The polymer template is subsequently removed by high-temperature oxidative thermal decomposition (calcination), and the remaining titanium dioxide skeleton is coated with silicon using chemical vapor deposition (CVD) at \(500^\circ\mathrm{C}\)~\cite{muller2013silicon}. Figure~\ref{fig:Fig1}~B depicts the final silicon network with a volume filling fraction of \(\phi \sim 20\text{--}25\,\%\), as estimated from focused ion beam (FIB) milling tomography. For this filling fraction, the peak of the density--density correlation function, corresponding to an effective lattice constant, lies at $a = 2.55 \bar d =2.04\,\mu\mathrm{m} $ (see Supplementary Materials, fig.~\ref{fig:FigS1}). In the experiments, as in previous studies, we observe small residual pockets or voids after silicon infiltration, lowering the effective rod refractive index. Staude \textit{et al.} reported an index reduction of about 10\% for silicon woodpile photonic crystals prepared by DLW templating, while Marichy \textit{et al.} reported a reduction of about 13\% for titanium dioxide inverse woodpile photonic crystals~\cite{staude2010fabrication,marichy2016high}. Consequently, we estimate the average material refractive index to be approximately 10--15\% lower than the bulk refractive index of dense or slightly porous amorphous silicon (\(n \sim 3.4 \pm 0.05\)) at the wavelengths considered here.

\subsection*{Optical characterization}
\begin{figure}[p]
  \centering
  \includegraphics[width=\textwidth]{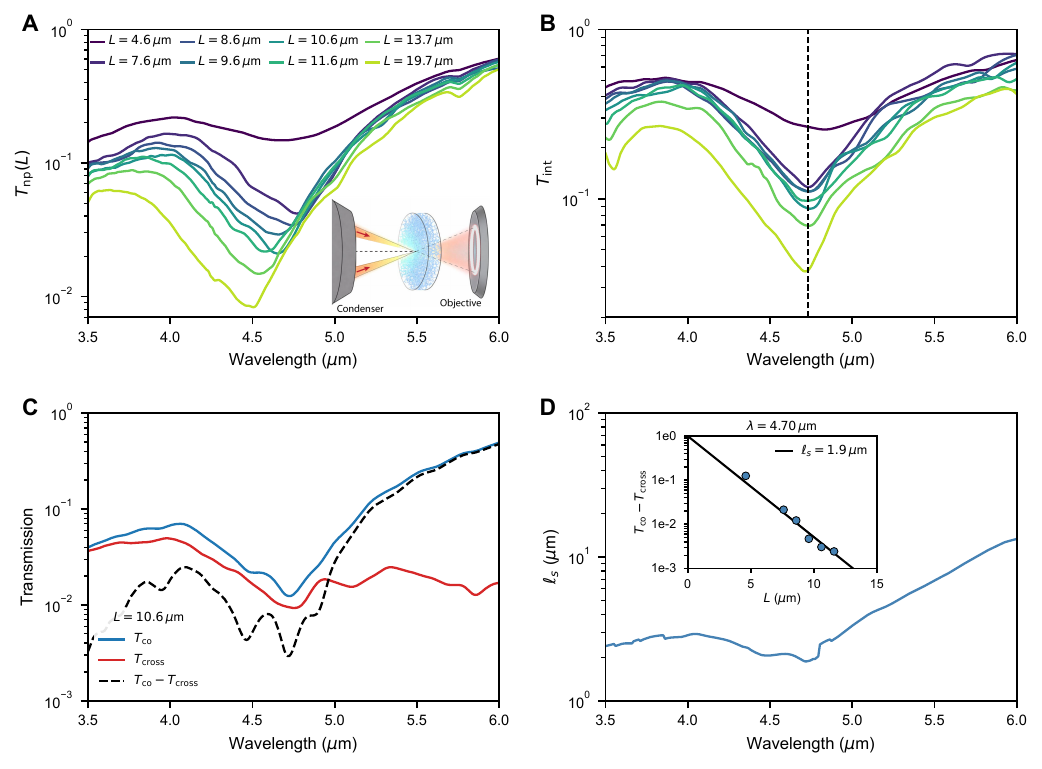}
  \caption{\textbf{Transmission properties of amorphous gyroid networks (AGNs).}
  \textbf{(A)} Unpolarized transmission spectra $T(\lambda)$ for different sample thicknesses $L$ ($\mu$m).
  \textit{Inset:} measurement geometry --- a 36X Cassegrain condenser with a numerical aperture ${\rm{NA}}=0.52$~\cite{newport_50102_02} and a matched objective collecting light transmitted through the sample~\cite{Aeby2021}.
  \textbf{(B)} Total integrated transmission $T_\mathrm{int}(\lambda)$, reconstructed from the measured
  polarization components to account for the diffuse flux outside the objective cone.
  \textbf{(C)} Co- and cross-polarized transmission and their difference (the ballistic component
  $T_\mathrm{co}-T_\mathrm{cross}$) for a representative sample ($L=10.6\,\mu$m).
  \textbf{(D)} Ballistic scattering mean free path $\ell_s(\lambda)$ extracted across the spectral range.
  The inset shows the ballistic component near the transmission minimum, which follows
  $T_b(L)=\exp(-L/\ell_s)$ (line) and yields $\ell_s=1.9\,\mu$m at $\lambda=4.7\,\mu$m.}
  \label{fig:Fig2}
\end{figure}

We evaluate the optical transport properties of the networks using a Fourier-transform infrared (FTIR) microscope equipped with two-mirror 36X Cassegrain objectives (NA=0.52, Obscuration 17\%, Newport Corporation, 2026~\cite{newport_50102_02}), as shown in the inset of Fig.~\ref{fig:Fig2}~A. Strikingly, we observe a pronounced transmission minimum that becomes deeper with increasing slab thickness. The total transmission within the gap reaches values as low as $10^{-2}$. At short wavelengths, multiple light scattering and diffusive transport dominate, resulting in low transmission. At wavelengths above the gap, the samples rapidly become transparent, as expected for our uniform designs~\cite{leseur2016high}. The gap center wavelength lies at $\lambda_{\mathrm{gap}} = 4.7\,\mu\mathrm{m}$ for thin samples and shifts slightly to $\lambda_{\mathrm{gap}} \sim 4.5\,\mu\mathrm{m}$ for larger $L$.
From electron microscopy, we observe a small thickness-dependent shrinkage of the structures~\cite{liu2017three}, see Supplementary Materials, fig.~\ref{fig:FigS2}, which becomes more pronounced for thicker samples and accounts for the corresponding blueshift of the transmission minimum, a conclusion supported by our simulations showing no spectral shift.  For the subsequent analysis, we therefore correct the wavelength scale, resulting in corrections of up to $\sim 4.5\%$ and bringing the minima into alignment at $\lambda_{\min} \sim 4.7\,\mu\mathrm{m}$. We note that much more significant shrinkage can be achieved by heating the polymeric templates to temperatures around \(450~^\circ\mathrm{C}\)~\cite{liu2019structural,aeby2022fabrication}; however, this was not attempted in the present study.

To obtain the deepest transmission minimum, we systematically increased the amount of deposited silicon until reaching an optimum, as shown in the Supplementary Materials, fig.~\ref{fig:FigS3}. The subsequent recovery of the transmission toward transparency at higher filling fractions provides strong evidence that absorption does not play a significant role in the observed transmission minimum over the spectral range considered.

In photonic crystals, light transport is governed by Bragg scattering from the periodic refractive-index modulation. In amorphous photonic structures, the situation is more complex: although long-range periodicity is absent, structural correlations and strong refractive-index contrast can still give rise to optical resonances and a stop band, while outside the gap multiple scattering plays a central role in the transport of light. For light of wavelength $\lambda$ incident on a dielectric slab of thickness $L$, infinite in the transverse directions, we define the integrated transmission $T_\mathrm{int}(L)$ and reflection $R_\mathrm{int}(L)$ as the fractions of the incident flux that are transmitted and reflected, respectively, such that $T_\mathrm{int}(L)+R_\mathrm{int}(L)=1$ in the absence of absorption. We distinguish this integrated transmission from the ballistic contribution $T_b(L)$, which accounts only for light transmitted in the direct propagation direction of the incident beam. The total integrated transmission can therefore be decomposed as $T_\mathrm{int}(L)=T_b(L)+T_d(L)$, where $T_d(L)$ denotes the diffuse, or non-ballistic, contribution arising from multiple scattering. Light propagation in disordered dielectric media is commonly described by classical transport theory as the combined effect of ballistic propagation and (multiple) scattering~\cite{ishimaru1978wave,vera1996diffusely}. A strong suppression of the photonic density of states is therefore identified by deviations of the scattered light from this classical transport picture.

To separate light passing straight through from light scattered many times, we resolve the co- and cross-polarized components, $T_{\rm{co}}(L)$ and $T_{\rm{cross}}(L)$, for linearly polarized incident light using an IR wire-grid polarizer (WP25M-IRA, Thorlabs, USA). From this, we compute the total integrated transmission reconstructed from the measured components. The objective collects only $\simeq$~22\% of the diffuse flux emitted into the full hemisphere, so the cross-polarized signal must be upscaled to account for the undetected fraction. Assuming the diffuse radiance is isotropic over the hemisphere (Lambertian emission), the total transmission is
$T_\mathrm{int}(L) = T_\mathrm{nopol} + \left(\frac{1}{0.22} - 1\right) \times 2\,T_\mathrm{cross}$, where $T_\mathrm{nopol}$ captures both the ballistic and scattered contributions (collected within the objective cone), and $2\,T_\mathrm{cross}$ estimates the total diffuse flux, scaled by $1/0.22$ to recover the fraction missed outside the cone. Total transmission spectra $T_\mathrm{int}(\lambda)$ for slabs with thicknesses ranging from \(L = 5~\mu\mathrm{m}\) to \(20~\mu\mathrm{m}\) are shown in Fig.~\ref{fig:Fig2}~B. A decomposition into the different polarization components is shown in Fig.~\ref{fig:Fig2}~C for one sample.

By subtracting the cross-polarized signal from the co-polarized signal, we obtain an accurate estimate of the ballistic transmission,
\begin{equation}
T_b(L) \simeq T_{\rm co}(L) - T_{\rm cross}(L).
\label{Eq:SubBall}
\end{equation}
Eq.~\ref{Eq:SubBall} relies on the rapid depolarization of multiply scattered light, whereas the ballistic component preserves the incident polarization. Its validity for the slab thicknesses considered here is confirmed by FDTD simulations, see Supplementary Materials, fig.~\ref{fig:FigS4}. From $T_b(L)=\exp(-L/\ell_s)$, we extract the wavelength-dependent extinction length $\ell_s(\lambda)$ shown in Fig.~\ref{fig:Fig2}~D; a direct comparison of the experimental $\ell_s(\lambda)$ with the corresponding FDTD result is provided in the Supplementary Materials, fig.~\ref{fig:FigS5}. In classical scattering, $\ell_s(\lambda)$ is referred to as the scattering mean free path, whereas in a photonic stop band it is often called the Bragg length~\cite{joannopoulos2008molding,hasan2018finite}. Because the ballistic component decays exponentially both inside and outside a bandgap, it cannot by itself distinguish between different transport regimes. As shown in the inset of Fig.~\ref{fig:Fig2}~D, a clear exponential extinction is observed, with a decay length $\ell_s \simeq 2\,\mu\mathrm{m}$ near the transmission minimum. This value is comparable to the lattice parameter $a \approx 2\,\mu\mathrm{m}$, consistent with the presence of a strong photonic stop band.

\subsection*{Model comparison}
\paragraph{Numerical simulations.} We used the same digital template as in the fabrication step, consisting of elliptical rods of length $d = 0.8~\mu\mathrm{m}$ with an aspect ratio of $2.5$, a refractive index of $n = 3.4$, and a rod filling fraction of $\phi\approx 0.237$, closely matching the experimental conditions.
To reproduce the subwavelength-sized voids that form within the rods during fabrication, we add air voids at a volume fraction of $\approx 20\%$ inside the rods, thereby reducing the silicon filling fraction. Consequently, the cross-sectional area of the voids is $20\%$ of the rod cross-sectional area, and the voids have the same aspect ratio as the rods. The statistics of the void sizes are detailed further below. Within an effective-medium picture, the introduction of air voids lowers the effective refractive index. Treating a rod as bulk silicon ($n=3.4$, $\varepsilon_{\rm Si}=11.56$) containing an air-void fraction of $\approx20\%$, the Bruggeman approximation gives an effective refractive index $n_{\rm rod}\simeq2.93$, while the Maxwell–Garnett approximation gives $n_{\rm rod}\simeq2.96$.

To quantitatively compare spectroscopy experiments and simulations, we employ a hardware optimized finite-difference-time-domain (FDTD) solver (Tidy3D, Flexcompute Inc., USA), which directly solves Maxwell's equations and faithfully reproduces the experimental illumination and detection geometry~\cite{Flexcompute2022,yamilov2023anderson}. We use clean-cut slabs with thicknesses ranging \(L = 2\text{--}15~\mu\mathrm{m}\). We illuminate the slab with a Gaussian-pulsed plane-wave source containing the frequencies of interest and apply periodic boundary conditions perpendicular to the direction of propagation to simulate an infinite periodic structure. At each end of the simulation cell, we place adiabatic absorbers to avoid the reentry of the transmitted and reflected fields. We extract the total integrated transmission, and using a diffraction monitor at the back of the structures, we decompose the transmitted field into ballistic, co-polarized, and cross-polarized parts.

\begin{figure}[p]
  \centering
  \includegraphics[width=\textwidth]{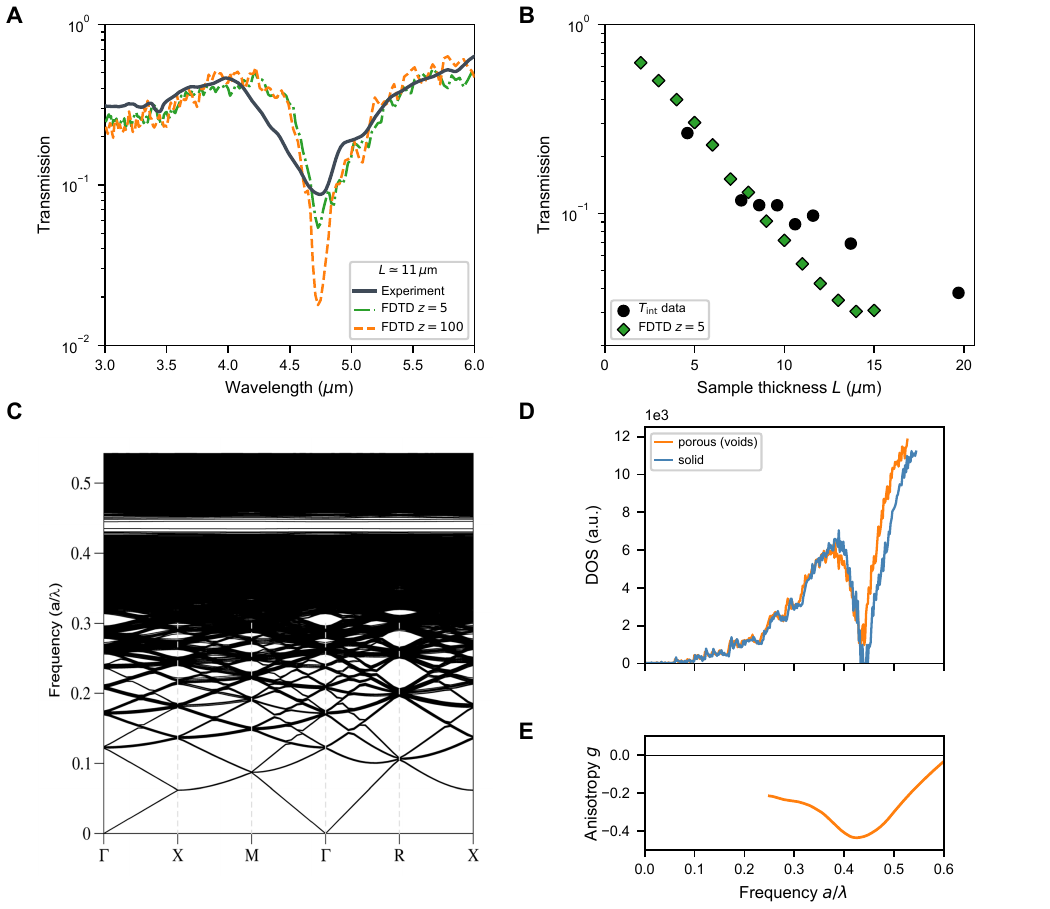}
  \caption{%
  \textbf{Experiment, FDTD, and band structure of the AGN photonic gap.}
  \textbf{(A)} Integrated transmission spectra $T_\mathrm{int}(\lambda)$ from the experiment ($L=10.6\,\mu\mathrm{m}$) and from FDTD simulations with 20\% porosity averaged over five
  realizations, for $L=11.0\,\mu\mathrm{m}$. Orange dashed line shows the results for monodisperse voids ($z=100$) and the green dash-dotted line the results for polydisperse voids ($z=5$).
  \textbf{(B)} Integrated transmission at $\lambda_\mathrm{gap}$ for different
  sample thicknesses $L$. Black full circles show $T_{\rm{int}}$ reconstructed from the polarization-resolved
  transmission spectra and compared with the FDTD polydisperse simulations (green diamonds).
  \textbf{(C)} Band diagram for an AGN composed of solid rods with refractive index $n=2.92$ in air and a filling fraction of $23.7\%$.
  \textbf{(D)} Density of states (DOS, orange line) averaged over the five realizations for polydisperse voids. For comparison,
  the blue line shows the DOS of the AGN modeled with homogeneous rods of
  refractive index $n=2.92$. \textbf{(E)} Scattering anisotropy parameter \(g = \langle \cos \theta \rangle\) derived from the scattering function of the network's digital template. %
}
  \label{fig:Fig3}
\end{figure}

The voids and inclusions seen in the FIB-SEM images after silicon infiltration (Fig.~\ref{fig:Fig1}~B) are highly heterogeneous: some rods are almost solid, whereas others contain large open regions. These inhomogeneities lower the effective rod refractive index and introduce additional structural disorder, thereby hindering the full opening of the otherwise narrow photonic band gap. We account for this by introducing disorder in the void volumes in the network model, using a Zimm--Schulz distribution while keeping the mean void fraction fixed at 20\%. Consequently, the effective rod index and the gap position remain independent of the degree of disorder. The void polydispersity index, given by $\mathrm{PDI}=1/(1+z)$, indicates that small $z$ corresponds to broad distributions, whereas large $z$ corresponds to nearly monodisperse voids. We perform FDTD simulations for $z=5$ ($\mathrm{PDI}=1/6$) and $z=100$ ($\mathrm{PDI}=1/101$), representing highly polydisperse and nearly monodisperse void distributions, respectively. Spectra are averaged over five independent realizations for each $z$. Comparison with the experimentally reconstructed total transmission in Fig.~\ref{fig:Fig3}~A and B shows excellent agreement for different slab thicknesses in the polydisperse case.

Using the supercell method with periodic boundary conditions to calculate the band structure, Fig.~\ref{fig:Fig3}~C, and the photonic density of states, Fig.~\ref{fig:Fig3}~D, we predict a full photonic band gap for homogeneous rods with $n \simeq 2.92$. The gap is centered at $a/\lambda_{\mathrm{gap}} \simeq 0.441$, corresponding to $\lambda_{\mathrm{gap}} \simeq 4.63~\mu\mathrm{m}$, and has a relative width of $\Delta\lambda/\lambda_{\mathrm{gap}} \simeq 3\%$~\cite{johnson2001block}. Thus, the band-structure calculations reproduce the gap position quantitatively. Introducing $20\%$ voids of equal size inside the rods with a refractive index of $n=3.4$ almost exactly reproduces the density of states of the lower index solid rods, as shown in the Supplementary Materials, fig.~\ref{fig:FigS6}. Introducing randomness in the void volumes (polydispersity with $z=5$), while keeping the total void volume constant, fills the gap with states, while $\lambda_{\mathrm{gap}}$ remains essentially unchanged. Although the gap closes, the overall density of states remains strongly reduced, as shown in Fig.~\ref{fig:Fig3}~D.

Additionally, in Fig.~\ref{fig:Fig3}~E we show the scattering anisotropy parameter defined as $g=\langle \cos \theta \rangle$, where $\theta$ denotes the scattering angle. We calculate the scattering anisotropy parameter \(g = \langle \cos \theta \rangle\) from the scattering function of the network, rescaling the wavelength using the effective refractive index of the material, see Materials and Methods. Similar to photonic crystals, the structural information encoded in $g(\lambda)$ determines where the gap opens in $k$-space. The effective refractive index $n_{\text{eff}}$ then maps this $k$-space position to a wavelength via the dispersion relation~\cite{sakoda2005optical}. In contrast, near-field and multiple-scattering corrections primarily influence the gap width rather than its central position. As shown in Fig.~\ref{fig:Fig3}~E, \(g\) becomes strongly negative, and its minimum aligns with the center of the bandgap, confirming the consistency of this approach. The values of $g$ obtained in the low-index limit are not expected to correspond exactly to the strongly scattering samples ($\ell_s\sim a$). We do, however, expect negative values around the gap: any deviation from the result shown in Fig.~\ref{fig:Fig3}~E should remain in the range $g\in[-0.75,-0.25]$, so the consequence for $\ell^\ast=\ell_s/(1-g)$ used later in the text is minor.

\begin{figure}[p]
  \centering
  \includegraphics[width=0.7\textwidth]{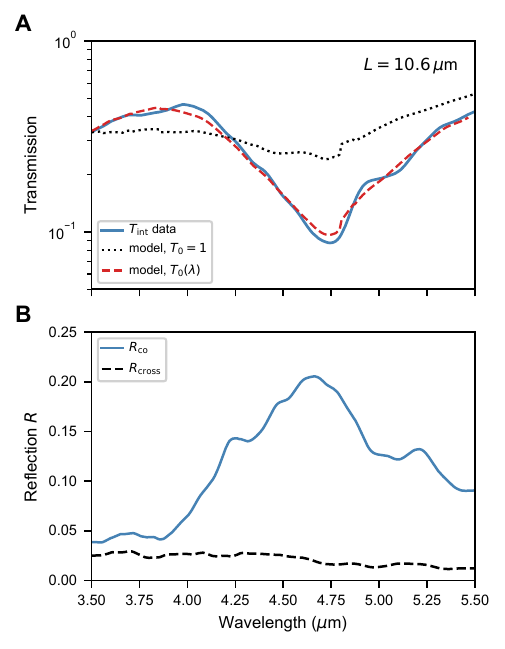}
  \caption{%
    \textbf{Transmission and polarized reflection}
    for the $L = 10.6\,\mu\mathrm{m}$ sample.
    \textbf{(A)} Integrated transmission $T_\mathrm{int}(\lambda)$ (solid line),
    compared with the diffuse-transport model
    $T_\mathrm{int}(\lambda) = e^{-L/\ell_s} + T_0\,b(L)$ evaluated with a fixed source
    amplitude $T_0 = 1$ (dotted) and with the fitted, wavelength-dependent
    amplitude $T_0(\lambda)$ (dashed). \textbf{(B)} Co- ($R_\mathrm{co}$) and cross-polarized ($R_\mathrm{cross}$)
    reflection of the same sample as measured.  }
  \label{fig:Fig4}
\end{figure}

\paragraph{Coherent, Bragg-like reflection.}
In Fig.~\ref{fig:Fig4}~A we replot the integrated-transmission spectrum for one sample ($L=10.6\,\mu$m) from Fig.~\ref{fig:Fig2}~B. Earlier work indicates that a reduced DOS leads to coherent reflection~\cite{scheffold2022transport}; the reflection spectra of the same sample directly confirm this, showing that the transmission minimum coincides with strong coherent reflection, Fig.~\ref{fig:Fig4}~B. Within the gap ($R_\mathrm{co}\simeq 0.2$ near $\lambda \approx 4.7\,\mu\mathrm{m}$), the co-polarized reflection peaks sharply, while the cross-polarized signal remains low and featureless ($R_\mathrm{cross}\simeq 0.02$). The reflection is therefore predominantly coherent and polarization-preserving, rather than depolarized by multiple scattering, and exhibits a Bragg-like response despite the absence of long-range periodic order. In the ideal limit of a vanishing density of states, all incident light is reflected coherently rather than coupling to diffusive multiple-scattering modes.

\paragraph{Modelling light transport.}
To make this point more quantitative, we compare the transmission spectra of slabs with different thicknesses to predictions from classical transport theory. Multiple light scattering theory assumes the existence of propagating optical modes that support diffusive transport, with the photonic density of states approaching the classical form $\rho_0(\nu)=\frac{8\pi \nu^2}{c^3}$, where $c$ denotes the speed of light in the effective medium. Near a complete or nearly complete photonic band gap, however, the density of available optical states is strongly suppressed, and this conventional diffuse-transport picture breaks down. Comparison with classical transport theory, therefore, provides a stringent test: if the measured transmission falls systematically below the prediction of classical multiple scattering, the attenuation cannot be explained by diffusive transport alone. To this end, we use the widely adopted Durian–Vera theory~\cite{vera1996diffusely}, which is an approximate solution of classical radiative transfer for a finite scattering slab. The model separates ballistic attenuation from multiply scattered diffuse transport $T_\mathrm{int}(L)=T_b(L)+T_d(L,g)$ with $T_d(L,g)\equiv b(L,g)$ in the classical limit and
\begin{equation}
b(L,g)
=
\frac{(1+z_0)(1-e^{-L/\ell_s}) - e^{-L/\ell_s}(1-g)L/\ell_s}
{2z_0 + (1-g)L/\ell_s}.\label{eq:DurianTS}
\end{equation}
Here, $z_0 \simeq 1.8$ is a boundary parameter related to surface reflectivity (see Materials and Methods). We estimate \(\ell^\ast = \ell_s/(1 - g)\) from the experimental values of \(\ell_s\) and the calculated values for the scattering anisotropy parameter $g$.

\begin{figure}[p]
  \centering
  \includegraphics[width=\textwidth]{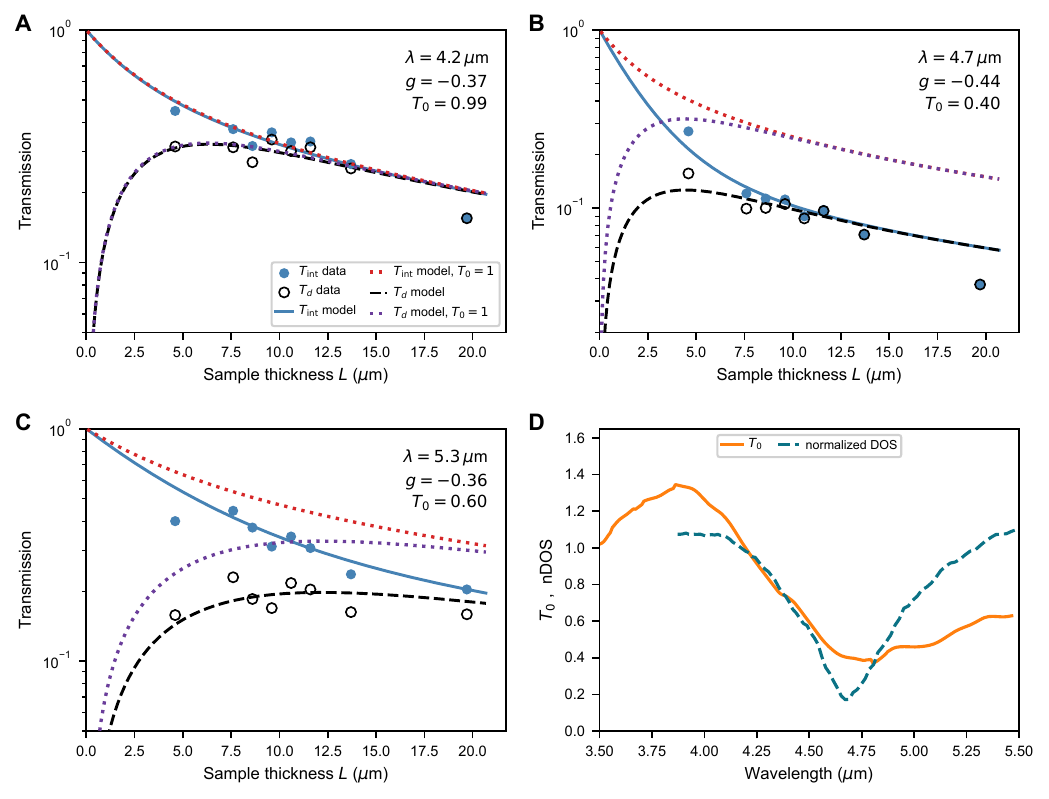}
  \caption{%
    \textbf{Integrated and diffuse transmission across the gap.}
    Total integrated transmission $T_\mathrm{int}(L)$ (filled symbols) and diffuse
    transmission $T_d(L) = 2\,T_\mathrm{cross}(L)/0.22$ (open symbols) as a function of
    sample thickness $L$, at three wavelengths:
    \textbf{(A)} $\lambda = 4.2\,\mu\mathrm{m}$,
    \textbf{(B)} $\lambda = 4.7\,\mu\mathrm{m}$ (band-gap centre), and
    \textbf{(C)} $\lambda = 5.3\,\mu\mathrm{m}$.
    Solid lines show the diffuse-transport model
    $T_\mathrm{int}(L) = e^{-L/\ell_s} + T_0\,b(L)$ and dashed lines the diffuse
    term alone $T_d(L) = T_0\,b(L)$, with the build-up factor
    $b(L) = \big[(1+z_0)(1-e^{-L/\ell_s}) - (1-g)\tfrac{L}{\ell_s}e^{-L/\ell_s}\big]
    /\big[2z_0 + (1-g)\tfrac{L}{\ell_s}\big]$ and $z_0 = 1.8$. At each wavelength
    the scattering mean free path $\ell_s$ and anisotropy $g$ are fixed from the
    spectral fits, and the single free parameter $T_0$ is obtained by fitting the
    diffuse component $T_d(L)$ alone (open symbols) across all sample
    thicknesses. The classical bound using $T_0\equiv 1$ are shown as dotted lines.
    \textbf{(D)} Fitted diffuse source amplitude $T_0(\lambda)$, which drops sharply across the gap. Its minimum coincides with the transmission minimum and with the maximum of the coherent co-polarized reflection (Fig.~\ref{fig:Fig4}~A,B). The dashed line shows the normalized density of states, obtained from the DOS (porous) shown in Fig.~\ref{fig:Fig3}~D by dividing out the trivial low-frequency Debye scaling, $(a/\lambda)^2$, of a homogeneous medium. %
  }
  \label{fig:Fig5}
\end{figure}

As shown in Fig.~\ref{fig:Fig5}~A, dotted lines, the model reproduces the transmission away from the gap region with no adjustable parameters. In contrast, pronounced deviations emerge around $\lambda = 4.7\,\mu\mathrm{m}$, Fig.~\ref{fig:Fig5}~B, where the transmission falls systematically below the classical bounds, partially recovering at larger wavelengths, Fig.~\ref{fig:Fig5}~C. To recover good agreement with the experimental data, we follow the suggestion of Ref.~\cite{scheffold2022transport} and multiply the diffuse term $b(L)$ of Eq.~\eqref{eq:DurianTS} by a prefactor $T_0$, modulating the diffuse light amplitude $T_d(\lambda)$. In the picture proposed earlier~\cite{scheffold2022transport}, $T_0$ accounts for the missing fraction of incident light that is directly reflected by the gap and therefore does not couple into diffusive transport. This suppression is consistent with a breakdown of conventional diffusive transport and a reduced photonic density of states, in agreement with the independent band-structure and density of states calculations.

The corresponding $T_0(\lambda)$, extracted from a joint fit to all samples for which scattering is sufficiently strong ($L_\mathrm{max}/\ell_s > 3$), is shown in Fig.~\ref{fig:Fig5}~D.
The spectral evolution of $T_0(\lambda)$, including its minimum within the gap and slight overshoot near the band edge, closely follows the independently calculated normalized DOS. At longer wavelengths, $T_0(\lambda)$ recovers slowly.
Above $5.5\,\mu$m wavelength, the sample rapidly becomes too transparent, and we can no longer apply the polarization-resolved analysis.

\subsection*{Anderson localization of light}
\begin{figure}[p]
  \centering
  \includegraphics[width=0.75\textwidth]{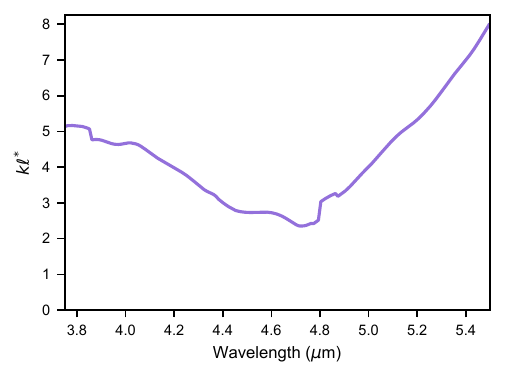}
\caption{\textbf{Ioffe--Regel parameter for Anderson localization}. The solid line shows the dimensionless transport mean free path, $k\ell^\ast = 2\pi \ell^\ast/\lambda_{\rm eff}$, using the values of $\ell_s$ shown in Fig.~\ref{fig:Fig2}~D and the scattering anisotropy parameter shown in Fig.~\ref{fig:Fig3}~E. For the wavelength in the effective medium, we use $\lambda_{\rm eff}=\lambda/n_{\rm eff}$ with $n_{\rm eff}\simeq 1.35$.}
  \label{fig:Fig6}
\end{figure}

A reduced but finite photonic density of states near a disorder-broadened band gap provides favorable conditions for Anderson localization of light, as originally proposed by John~\cite{john1987strong}. Localization is distinct from bandgap formation: its hallmark is an enhanced return probability of scattered waves, so that transport ceases and the transmission decays over a localization length $\xi > \ell_s$. It is notoriously difficult to reach in three dimensions for vector (electromagnetic) waves~\cite{Sperling2016,skipetrov2016red}, whereas it is met more readily for scalar waves or in one- and two-dimensional systems~\cite{hu2008localization,yamilov2014position,froufe2017band,Meek2026Localization,Granchi2026Spectral}. The localization transition in three dimensions is generally predicted for $k\ell^\ast$ on the order of unity, known as the Ioffe--Regel criterion, although the exact value is unknown and may be system dependent~\cite{skipetrov2018ioffe}. A reduced but finite density of states may raise the return probability~\cite{monsarrat2022pseudogap} and shift the transition to larger $k\ell^\ast$. Recent simulations report localization of vector waves in three-dimensional networks similar to ours, as well as in disordered metals~\cite{haberko2020transition,scheffold2022transport,yamilov2025anderson,goicoechea2026experimental}. For our networks, the Ioffe--Regel parameter $k\ell^\ast = 2\pi\ell^\ast/\lambda_{\rm eff}$, with $\ell^\ast=\ell_s/(1-g)$ and $\lambda_{\rm eff}=\lambda/n_{\rm eff}$, drops to $k\ell^\ast \simeq 2$ in the stop-gap region around $\lambda \simeq 4.7~\mu\mathrm{m}$, Fig.~\ref{fig:Fig6}, concurrent with the suppressed density of states and indicating proximity to the localization onset. This suggests our systems may approach, or possibly have already entered, Anderson localization. However, with the present experimental tools and finite sample thicknesses, we cannot yet unambiguously demonstrate Anderson localization of electromagnetic waves in lossless dielectrics; establishing such signatures remains an important objective for future work.

\section*{DISCUSSION}
The central result of this work is the experimental identification of coherent, Bragg-like reflection and an associated strong suppression of the photonic density of states in a three-dimensional disordered dielectric network, established through polarization-resolved transmission and reflection measurements and supported by band-structure calculations and FDTD simulations. Further improvements in fabrication should enhance the decay of the non-ballistic transmission within the gap, approaching the full bandgap behavior predicted for the idealized structure. Introducing polarization-resolved spectroscopy and demonstrating non-classical light transport opens new opportunities to explore other disorder-induced phenomena, in particular strong Anderson localization (SAL) of light~\cite{anderson1985question,lagendijk2009fifty,yamilov2023anderson}. Numerical studies indicate that dielectric networks supporting photonic bandgaps can promote the emergence of strong Anderson localization near the bandgap regime~\cite{haberko2020transition}. These phenomena should become experimentally accessible in samples with thicknesses much larger than the localization length, \(\xi \sim 5a \)~\cite{scheffold2022transport}.

\section*{MATERIALS AND METHODS}

\subsection*{Fabrication details}
We fabricate AGN polymer templates using direct laser writing (DLW) lithography with IP-Dip photoresist (Nanoscribe, Germany) and $n_\text{IP-Dip}=1.53$. Templates are written in a calcium fluoride optical window, which is transparent in the mid-infrared regime. The printed structures have a circular footprint with a diameter of $126\:\mu \mathrm{m}$. We use galvo writing mode with a speed of $15000\,\mu\mathrm{m}/\mathrm{s}$ and nominal laser power $21\,\mathrm{mW}$. The fabricated structures are then immersed in PGMEA (propylene glycol monomethyl ether acetate) to remove the unpolymerized photoresist and dried by means of critical-point drying. During laser printing, the material undergoes a density change as liquid monomers transform into a solid polymer network. This typically causes a small volumetric reduction, often reported at less than 2\%. Previously, an additional reduction of a few percent was reported to occur during the solvent development and drying stages~\cite{liu2017three}.

To increase the refractive index of the structures, we coat the polymer template with a thin layer of titanium dioxide by atomic layer deposition ($\sim$10\,nm) at a moderate temperature ($\sim$130$^\circ$C), which contributes approximately 1\% to the volume filling fraction. The polymer template is subsequently removed by calcination in an oxygen atmosphere at 400$^\circ$C. Earlier work has shown that the polymer decomposes under these conditions, as confirmed by thermogravimetric analysis (TGA) \cite{muller2013silicon}. This titanium dioxide-stabilized structure is then coated with silicon using chemical vapor deposition (CVD) at 500$^\circ$C. To achieve optimal accuracy and structural uniformity, we target a band gap center wavelength in the mid-infrared, within the range $4\,\mu\mathrm{m} < \lambda_\text{Gap} < 5\,\mu\mathrm{m}$, where typical infrared absorption bands are absent, and the refractive index of silicon is $n \sim 3.4 \pm 0.05$~\cite{salzberg1957infrared}.

\subsection*{FIB-SEM analysis}
The structures were characterized using scanning electron microscopy (SEM) and focused ion beam (FIB) milling. To minimize charging, they were sputter-coated with a thin layer of gold ($\sim 4$\,nm) and subsequently imaged using a Mira3 LM FE scanning electron microscope (Tescan). The cross-section shown in Fig.~\ref{fig:Fig1} was prepared using a Scios~2 focused ion beam scanning electron microscope (FIB--SEM, Thermo Fisher). Before milling, the sample surface was coated with a protective platinum layer by ion-beam-induced deposition.

\subsection*{MPB supercell}
Applying the MIT Photonic Bands (MPB) supercell method~\cite{johnson2001block} for our system parameters, $\bar d = 0.8~\mu$m, $n = 2.92$, and a rod aspect ratio of $2.5$, predicts a narrow full gap in the photonic density of states with a gap center wavelength of $\lambda_{\mathrm{gap}} = 4.63~\mu$m and a relative width of $\Delta \lambda / \lambda_{\mathrm{gap}} \sim 3\%$. Introducing voids with $z = 5$ fills the gap with states. To isolate the gap-induced suppression, we divide out the trivial low-frequency (Debye) scaling of a homogeneous effective medium, $\mathrm{DOS} \propto (a/\lambda)^2$. Starting from the simulated $\mathrm{DOS}(a/\lambda)$
of the voided network (Fig.~\ref{fig:Fig3}~D, orange curve), we fit the
low-frequency behavior $\mathrm{DOS} = C\,(a/\lambda)^2$ over $a/\lambda < 0.4$
with a single free parameter $C$ forced through the origin and define the
normalized density of states as
$\mathrm{nDOS} = \mathrm{DOS}\,/\,[\,C\,(a/\lambda)^2\,]$. The reduced frequency
$u \equiv a/\lambda$ is mapped to wavelength through $\lambda = a/u$ with
$a = 2.04\,\mu\mathrm{m}$.
By construction, $\mathrm{nDOS} \to 1$ is in the low-frequency effective-medium regime,
so the pronounced dip across the gap directly measures the fractional suppression
of the photonic DOS, as shown in Fig.~\ref{fig:Fig5}~D.

\subsection*{Computation of the anisotropy factor $g$ from the scattering function}
We numerically calculate the anisotropy factor $g=\langle \text{cos}\ \theta \rangle$ operating in the Rayleigh–Gans–Debye (RGD) single-scattering regime \cite{Debye1949,Bohren1983,vandeHulst1957}, where the differential scattering cross-section ($\frac{d\sigma}{d\Omega}(\theta,\lambda)$) depends entirely on the structure factor ($S(q)$) of the permittivity distribution of the structure.
In the RGD approximation, the differential cross section per unit solid angle is defined as:
\begin{equation}
    \frac{d\sigma}{d\Omega}(\theta,\lambda)\propto k^4(1+\text{cos}^2\ \theta)/2\ S(q(\theta))
\end{equation}
where $S(q)$ is the spherically averaged 3D FFT power spectrum of the mean-subtracted permittivity contrast, obtained from the digital template used for both simulations and experiments, integrated over $\theta \in (0,\pi)$. We can then intensity-weight average the cosine of the scattering angle:
\begin{equation}
    g=\langle \text{cos}\ \theta \rangle = \frac{\int\frac{d\sigma}{d\Omega}\ \text{cos}\ \theta\ \text{sin}\ \theta\ d\theta}{\int\frac{d\sigma}{d\Omega}\ \text{sin}\ \theta\ d\theta}
\end{equation}
If $g=0$ the scattering is isotropic, if $g<0$ most scattering is in the backward direction relative to propagation, and for $g>0$ we are in the forward scattering regime.

\subsection*{Multiple scattering and photon diffusion model}
We determine the ballistic component \(T_b(L)\) by subtracting the cross-polarized
signal from the co-polarized transmitted signal (Eq.~\ref{Eq:SubBall}). This approach
assumes that the ballistic component retains the initial polarization, whereas the
scattered contribution \(T_d(L)\) becomes randomly polarized, such that
\(T_{\mathrm{co}}(L) = T_b(L) + T_d(L)/2\) and \(T_{\mathrm{cross}}(L) = T_d(L)/2\).
This relation is exact only once the scattered light is fully randomized and transport
is diffusive (\(L/\ell_s \gg 1\)). For smaller thicknesses, the transmitted signal
retains partial memory of the initial polarization; therefore,
\(T_{\mathrm{co}}(L) - T_b(L) > T_{\mathrm{cross}}(L)\). We set the threshold for applying the polarization-resolved analysis to
\(L > 3\,\ell_s\), noting that in our case \(g < 0\) and \(\ell^\ast < \ell_s\), which
is a more favorable situation than for positive \(g\)~\cite{rojas2004depolarization}.
Two factors favor the accuracy of our approach to measure \(T_b(L)\). First, the Cassegrain objective collects essentially the entire ballistic component, but only about one fifth of the diffusely scattered contribution. Second, in and near the gap, the diffuse transmission is strongly suppressed. Consequently, the diffuse signal does not dominate the ballistic component to the extent that would be expected for conventional diffusive light transport.

We therefore analyze polarization-resolved transmission spectra to disentangle the
different contributions to \(T_\mathrm{int}(L)\). The illumination comes up through the CaF$_2$ substrate, hits the sample, which forward-scatters into air. The co-polarized transmission measured with the
FTIR objective, \(T_{\mathrm{co}}(L)\), contains the collinearly propagating ballistic
beam as well as a fraction of scattered light \(T_d(L)\) collected by the objective. By contrast, the
cross-polarized transmission contains only half the diffusively scattered contribution,
\begin{equation}
T_{\mathrm{cross}}(L) \simeq 0.22\, T_d(L)/2. \label{eq:TCross}
\end{equation}
Here the Cassegrain objective collects light within a numerical aperture of ${\rm{NA}}=0.52$ and an obscuration of 17\%. For a Lambertian diffuse emitter, the corresponding
collected fraction is $\mathrm{NA}^2(1-0.17)=0.22$ of the scattered intensity. The actual value may differ slightly which would shift the $T_0(\lambda)$ curves proportionally but would not qualitatively affect our analysis, as shown in the Supplementary Materials, fig.~\ref{fig:FigS7}. We note that the supplier of the objective also specifies an angular collection range from $15^\circ$ to $30^\circ$, which would correspond to a factor of $0.183$. In the present study, we use the likely more accurate values of the numerical aperture and obscuration provided in the same data sheet~\cite{newport_50102_02}.

The total integrated transmission is
\(T_\mathrm{int}(L) = T_b(L) + T_d(L) \simeq T_b(L) + \tfrac{1}{0.22} \times 2\,T_{\mathrm{cross}}(L)\),
or equivalently
\(T_\mathrm{int}(L) = T_\mathrm{nopol}(L) + \left(\tfrac{1}{0.22} - 1\right) \times 2\,T_{\mathrm{cross}}(L)\),
where \(T_\mathrm{nopol}(L) = T_b(L) + 2 T_{\mathrm{cross}}(L)\) captures
both the ballistic and scattered contributions (collected within the objective cone)
and \(2\,T_{\mathrm{cross}}(L)\) estimates the total diffuse flux, scaled by \(1/0.22\)
to recover the fraction missed outside the cone.

In photon diffusion theory, the extrapolation-length coefficient \(z_0\) at a planar boundary accounts for partial internal reflection, which depends on the refractive index contrast between the sample and its surroundings.
Using the 3D Bruggeman effective-medium approximation in terms of permittivities,
with \(n\simeq 3 \Rightarrow \varepsilon_{1}\simeq 9\), host \(n_h=1 \Rightarrow \varepsilon_{2}=1\), and \(\phi \simeq 0.237\), we obtain
\(\varepsilon_{\rm eff}=1.82\) and \(n_{\rm eff}=\sqrt{\varepsilon_{\rm eff}} \approx 1.35 \pm 0.04\). We adopt the Bruggeman effective index throughout. For our filling fraction and index contrast, the Maxwell--Garnett estimate (silicon inclusions in air) gives \(n_{\rm eff}\simeq1.27\).
For \(n_{\rm eff} \approx 1.35 \pm 0.04\) with symmetric air boundaries, we obtain \(z_0 \approx 1.8 \pm 0.15\). We note that the sample rests on a CaF$_2$ substrate, which reduces internal reflection; however, silicon deposited on the substrate increases the reflectivity of the interface. For simplicity, we therefore use the constant value \(z_0 = 1.8\) throughout.

\section*{Supplementary Materials}
Supplementary Text and Fig.~S1 to Fig.~S7 are included in this document, following the
reference list below.

\section*{Acknowledgments}

\paragraph{Funding.} This work was supported by the Swiss National Science Foundation (SNSF) through Sinergia Grant No. 216659. A.A.U. and F.S. acknowledge support from SNSF Grant No. 188494, and L.S.F.P. from SNSF Grant No. 197146. M.F. acknowledges support from the EPSRC (United Kingdom) under Grant No. EP/Y016440/1 and EP/Y016440/2 awards. We thank Aaron Shih, Ullrich Steiner, and Dave Pine for discussions.

\paragraph{Author contributions.} A.A.U. implemented the digital templates, carried out all the experiment and processed the data, with contributions from G.A. F.H.A. carried out the FDTD simulations. F.H.A. also calculated the anisotropy factor in collaboration with M.R. and A.A.U. L.S.F.-P. and F.S. supervised this numerical part of the project. M.F. provided the LSU digital templates and calculated the band diagram and density of states. F.S. supervised the project, led the data analysis, and drafted the manuscript. All authors contributed to the analysis of the data and the writing of the manuscript.

\paragraph{Competing interests.} The authors declare that they have no competing interests.

\paragraph{Data and materials availability.} All experimental data and numerical output shown in the paper and/or the Supplementary Materials will be uploaded to the repository Zenodo. Additional data related to this paper may be requested from the corresponding author.

\paragraph{Generative AI tools.} Generative AI tools were used solely for language editing (grammar and phrasing) and plotting. The authors take full responsibility for the scientific content of the manuscript.

\section*{REFERENCES AND NOTES}
\bibliography{LSUBib2026b_MF}

\clearpage
\section*{FIGURES}

\clearpage

\clearpage

\clearpage

\clearpage

\clearpage

\clearpage
\section*{SUPPLEMENTARY MATERIALS}
\setcounter{section}{0}
\setcounter{figure}{0}
\setcounter{table}{0}
\setcounter{equation}{0}
\renewcommand{\thefigure}{S\arabic{figure}}
\renewcommand{\thetable}{S\arabic{table}}
\renewcommand{\theequation}{S\arabic{equation}}
\renewcommand{\theHfigure}{S\arabic{figure}}
\renewcommand{\theHtable}{S\arabic{table}}
\renewcommand{\theHequation}{S\arabic{equation}}

\section*{Volume filling fraction and density--density correlation function $C(\Delta r)$}
\begin{figure}[p]
\centering
\includegraphics[width=.6\textwidth]{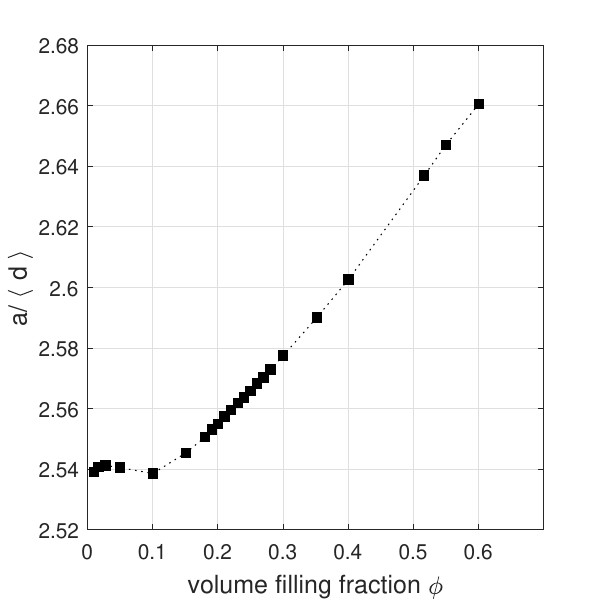}
\caption{ $a/ \langle d \rangle$ as a function of $\phi$.
}
\label{fig:FigS1}
\end{figure}

The characteristic length $a$ is defined as the second peak of the density–density correlation function $C(\Delta r)$:
\begin{equation} \label{cr}
    C(\Delta r)=\frac{ \langle \rho(r) \rho(r+\Delta r) \rangle - \langle \rho(r)  \rangle^{2}}{\langle \rho(r)  \rangle^{2}  }.
\end{equation}
Here, $\rho(r)$ is the three-dimensional material-density function and $ \langle \; \rangle $ is a spatial average. We use periodic digital representations of LSU networks consisting of arrays of $256 \times 256 \times256$. To compute Eq.~\ref{cr}, we use the Wiener–Khinchin theorem, which provides a relation between the autocorrelation in 3D and the power spectral density:
\begin{equation} \label{Wiener}
    C(\Delta r)\propto \mathcal{F}^{-1}(| \tilde{\rho} (k) |^2).
\end{equation}
$\tilde{\rho} (k)=\mathcal{F}(\rho (r))$ is the density in Fourier space. The value of $a$ depends on the volume filling fraction of the digital model; fig.~\ref{fig:FigS1} shows the value of $a/ \langle d \rangle$ as a function of $\phi$.

\section*{Residual shrinkage and the spectral shift of the transmission minimum}
\begin{figure}[p]
\centering
\includegraphics[width=.75\textwidth]{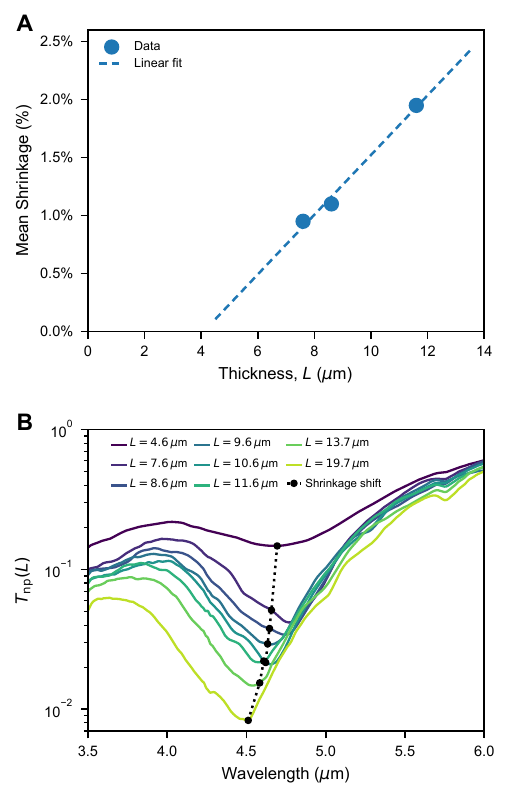}
\caption{
\textbf{(A)} Effective shrinkage used to estimate the structure, taken as one half of the mean shrinkage measured at the top using SEM, for sample thicknesses $L=7.6$, 8.6, and 11.6~$\mu$m. The dashed line shows a linear fit to the data,
$\mathrm{shrinkage}=(0.258~\%/\mu\mathrm{m})L-1.055~\%$.
\textbf{(B)} Unpolarized transmission spectra, $T_{\mathrm{nopol}}$, reproduced from Fig.~\ref{fig:Fig2}. The dotted line indicates the expected position of the transmission minimum based on the effective thickness-dependent shrinkage,
$\lambda_{\mathrm{gap}}\simeq 4.7\,[1-\mathrm{shrinkage}(L)/100]~\mu\mathrm{m}$.
}
\label{fig:FigS2}
\end{figure}

The position of the transmission minimum varies with slab thickness, as shown in fig.~\ref{fig:FigS2}. For the thinnest slabs, the minimum is located at $\lambda_{\mathrm{gap}}\approx 4.7~\mu\mathrm{m}$ and progressively blueshifts to $\lambda_{\mathrm{gap}}\approx 4.5~\mu\mathrm{m}$ as the slab thickness increases. This effect can be explained by a slight increase in sample shrinkage with increasing slab thickness.

To estimate this effect, we analysed SEM images of representative samples with nominal thicknesses of $8.4~\mu\mathrm{m}$, $9.4~\mu\mathrm{m}$, and $12.4~\mu\mathrm{m}$. By measuring the diameter at the top of each slab and comparing it with the nominal diameter of $126~\mu\mathrm{m}$, we estimate linear shrinkages of $1.9\%$, $2.2\%$, and $3.9\%$, respectively. This trend is consistent with the shift of $\lambda_{\mathrm{gap}}$ towards shorter wavelengths.

However, these values may somewhat overestimate the effective shrinkage of the entire slab because the lower layers, which are attached to the substrate, are expected to shrink less than the top surface. We therefore approximate the effective shrinkage relevant to the optical spectra as one half of the shrinkage measured at the top by SEM, as shown in fig.~\ref{fig:FigS2}~A. A linear fit yields the thickness-dependent shrinkage, $\mathrm{shrinkage}(L)$, which we use to estimate the corresponding spectral shift according to
$\lambda_{\mathrm{gap}}(L)\simeq 4.7\left[1-\frac{\mathrm{shrinkage}(L)}{100}\right]~\mu\mathrm{m}.$ The estimated shift due to shrinkage agrees well with the experimentally observed spectral shift, as shown in fig.~\ref{fig:FigS2}~B.

\section*{Evolution of the transmission spectra with the degree of silicon infiltration}

\begin{figure}[p]
\centering
\includegraphics[width=.9\textwidth]{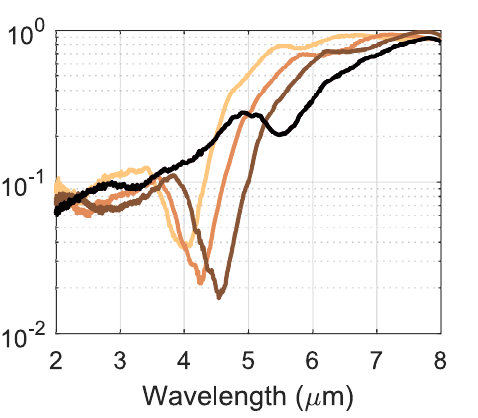}
\caption{Transmission spectra for a slab of thickness ($L=11.6\,\mu\mathrm{m}$) with increasing amounts of deposited silicon. Spectra corresponding to low silicon deposition are shown in light brown, whereas progressively darker colors indicate increasing amounts of deposited silicon. The third spectrum corresponds to the sample analyzed in the main manuscript. }
\label{fig:FigS3}
\end{figure}

To optimize the sample composition and evaluate the influence of absorption, we studied the evolution of the transmission spectra with increasing silicon infiltration. A representative example is shown in fig.~\ref{fig:FigS3} for a slab of thickness $L=11.6\,\mu\mathrm{m}$. Initially, the transmission minimum becomes progressively deeper. Once the optimal degree of infiltration is exceeded, however, the transmission minimum collapses, and the transmission increases again. Throughout the infiltration process, the position of the transmission minimum shifts continuously toward longer wavelengths. The numerical band-structure and transmission calculations indicate an optimal silicon filling fraction of approximately 20\text{--}30\%. The experimental results, therefore, confirm that selecting the sample with the deepest transmission minimum also yields a volume filling fraction within this range, consistent with the FIB--SEM analysis and the detailed numerical comparison presented in the main manuscript.
The non-monotonic evolution of the transmission provides further evidence that absorption does not play a significant role in our samples. The transmission initially decreases with increasing silicon deposition but rises again across the entire spectral range once substantially more silicon is added. By contrast, a transmission minimum caused primarily by infrared absorption would be expected to deepen systematically as the amount of absorbing material increases, as we observed for absorption features in other spectral ranges.

\section*{Validation of the polarization-resolved analysis}
\begin{figure}[p]
\centering
\includegraphics[width=.9\textwidth]{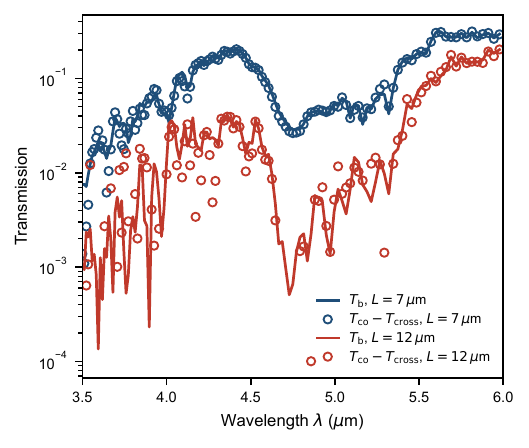}
\caption{Validation of Eq.~\ref{Eq:SubBall} using FDTD simulations (polydisperse case) and with a finite detection cone, mimicking the Cassegrain objective used in the experimental study. The ballistic transmission obtained directly from the simulations, $T_b(L)$, is compared with the estimate $T_{\mathrm{co}}(L)-T_{\mathrm{cross}}(L)$. Good agreement is observed over the investigated wavelength range and for the different slab thicknesses. When $T_{\mathrm{co}}(L)-T_{\mathrm{cross}}(L)$ becomes very small or even slightly negative, the values scatter around the noise floor. In Fig.~\ref{fig:Fig2} we therefore restrict the analysis to values $T_{\mathrm{co}}(L)-T_{\mathrm{cross}}(L) > 0.002$.}
\label{fig:FigS4}
\end{figure}

We determine the ballistic component, $T_b(L)$, by subtracting the cross-polarized transmitted signal from the co-polarized signal [Eq.~\ref{Eq:SubBall}]. This procedure assumes that the ballistic component retains its initial polarization, whereas the multiply scattered contribution, $T_d(L)$, becomes depolarized and is therefore equally distributed between the co- and cross-polarized channels:
\[
T_{\mathrm{co}}(L)=T_b(L)+\frac{T_d(L)}{2},
\qquad
T_{\mathrm{cross}}(L)=\frac{T_d(L)}{2}.
\]
For optically thin samples, the scattered transmitted light may retain a partial memory of the initial polarization. We therefore restrict the polarization-resolved analysis to $L>3\ell_s$. We note that, in our case, $g<0$ and hence $\ell^\ast<\ell_s$.
To validate this procedure, we also apply it to the FDTD simulation data, for which we have direct access to both $T_b(\lambda,L)$ and the polarization-resolved components $T_{\mathrm{co}}(\lambda,L)$ and $T_{\mathrm{cross}}(\lambda,L)$. The comparison is shown in fig.~\ref{fig:FigS4}. We find good agreement over the entire wavelength range and for the different slab thicknesses investigated.

\section*{Comparison of $\ell_s(\lambda)$ between FDTD and experiment}

\begin{figure}[p]
\centering
\includegraphics[width=.6\textwidth]{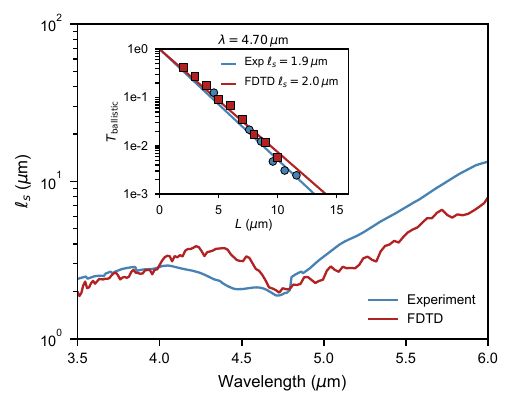}
\caption{\textbf{Scattering mean free path $\ell_s(\lambda)$.}
Spectral dependence of the scattering mean free path extracted from the
ballistic transmission component, comparing experiment (blue) and FDTD
simulation (red; filling fraction $0.237$). For
each wavelength, $\ell_s$ is obtained from a linear fit of
$\ln T_\mathrm{b}=-L/\ell_s$ versus sample thickness $L$ constrained through the
origin using thicknesses
$L = 4.6$--$19.7\,\mu\mathrm{m}$ (experiment, from
$T_\mathrm{co} - T_\mathrm{cross}$) and $L = 2$--$15\,\mu\mathrm{m}$
(FDTD, $T_\mathrm{ballistic}$). Only points with
$T_\mathrm{b} > 2\times10^{-3}$ are included in each fit, to exclude
values at or below the noise floor. \textit{Inset:} Representative ballistic-decay
fits at $\lambda = 4.7\,\mu\mathrm{m}$, yielding
$\ell_s = 1.9\,\mu\mathrm{m}$ (experiment) and
$\ell_s = 2.0\,\mu\mathrm{m}$ (FDTD). }
\label{fig:FigS5}
\end{figure}

Here, we compare the scattering mean free path extracted from the experimentally measured ballistic attenuation with that obtained from the attenuation of the directly transmitted beam in the FDTD simulations, as shown in fig.~\ref{fig:FigS5}.

\section*{Density of states for porous rods and lower-index solid rods}

\begin{figure}[p]
\centering
\includegraphics[width=.6\textwidth]{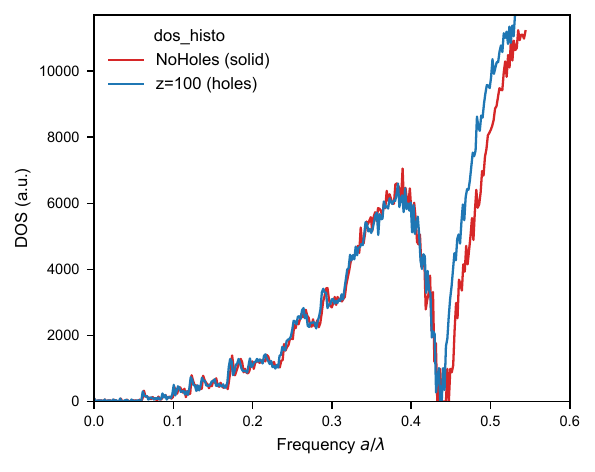}
\caption{\textbf{Density of states for homogeneous and porous rods.} The red line shows the density of states for a system composed of homogeneous rods with refractive index $n=2.92$, while the blue line shows the corresponding results for porous silicon rods ($n=3.4$) containing 20\% equally sized voids.}
\label{fig:FigS6}
\end{figure}

In fig.~\ref{fig:FigS6}, we compare the density of states obtained from MPB calculations for the same structures composed either of homogeneous rods with a lower refractive index or of a high-index material containing 20\% air voids. The resulting density of states is nearly identical in the two cases. The equally sized voids lead to a slight filling of the gap, but this effect is much weaker than in the case of polydisperse void volumes. The small deviation is likely related to the finite size of the voids, with smaller voids expected to reproduce the homogeneous-rod result more accurately. Since this case is not directly relevant to our study, where a substantially stronger filling of the gap is observed, we did not investigate this effect further.

\section*{Fraction of scattered light collected by the objective}
\begin{figure}[p]
\centering
\includegraphics[width=.6\textwidth]{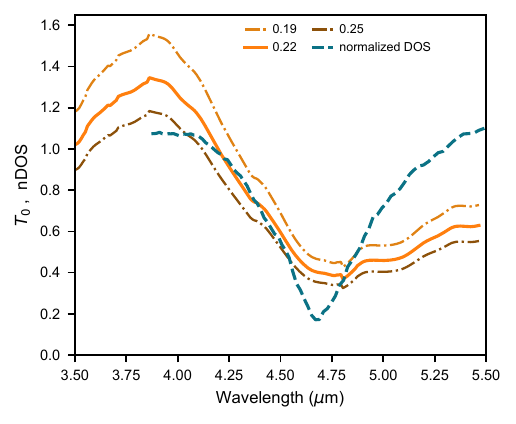}
\caption{\textbf{Dependence of $T_0$ on the detection cone.} The numbers indicate the fraction of diffuse light emitted by a Lambertian diffuser and collected by the FTIR microscope objective, shown for $0.19$, $0.22$ (the value used throughout), and $0.25$.}
\label{fig:FigS7}
\end{figure}

In the Materials and Methods section, we estimated the fraction of scattered light collected by the objective from the numerical aperture and central obscuration of the Cassegrain objective used in this study. The resulting value, 0.22, is an estimate, and the actual value may differ slightly, for example, due to refraction at the interfaces between the effective medium, the CaF$_2$ substrate, and air. In practice, the extracted values of $T_0$ scale inversely with the assumed collection fraction. Small variations within a reasonable range, $0.19$--$0.25$ (i.e. $0.22\pm0.03$), shift the $T_0(\lambda)$ curves vertically but do not qualitatively affect our analysis, as shown in fig.~\ref{fig:FigS7}.

\clearpage

\clearpage

\clearpage

\clearpage

\clearpage

\clearpage

\clearpage

\end{document}